\documentclass{elsart}
\usepackage{amssymb}
\usepackage{amsmath}
\usepackage{bm}

\newtheorem{definition}{Definition}
\newtheorem{stat}{Statement}

\journal{\textbf{Elsevier }   \qquad\qquad\qquad\qquad\qquad\qquad\qquad\qquad\qquad\qquad\qquad\qquad\qquad\qquad\qquad\qquad\qquad\qquad\qquad\qquad\qquad\qquad\qquad\qquad}

\begin{document}

\begin{frontmatter}
International Journal of Non-Linear Mechanics 47 (2012) pp. 688-693\\

\title{A consistent thermodynamical model \\ of incompressible media as limit case \\ of quasi-thermal-incompressible materials}

\author{Henri Gouin $^{*}$ \ \and\  Tommaso Ruggeri $^{ \dag}$}

\ead{henri.gouin@univ-amu.fr}
 \address { $^{*}$  M2P2, C.N.R.S. U.M.R. 7340 \&
 University of Aix-Marseille, \\ Case 322, Av. Escadrille
 Normandie-Niemen, 13397 Marseille Cedex 20 France.}

\address{$^{\dag}$  Department of Mathematics \& Research Center of Applied Mathematics, \\
University of Bologna, Via Saragozza 8, 40123 Bologna, Italy.}
\ead{tommaso.ruggeri@unibo.it}

\begin{abstract}
 In this paper we extend the conditions on
\emph{quasi-thermal-incompressible materials} presented in \cite{GMR} so that
they satisfy all the principles of thermodynamics, including the
stability condition associated with the concavity of the chemical
potential. We analyze the approximations under which a quasi-thermal-incompressible
medium can be considered as incompressible. We find that the pressure
cannot exceed a very large critical value and that the compressibility
factor must be greater than a lower limit that is very small.
The analysis is first done for the case of  fluids and then extended to the case of  thermoelastic solids.
\end{abstract}
\begin{keyword} Incompressible fluids and solids; Entropy principle;
Chemical potential concavity.
\PACS  47.10.ab; 62.20.D-; 62.50.-p; 64.70.qd; 65.40.De
\MSC  74F05; 74A10; 74A15; 76Bxx; 76E19
\end{keyword}

\end{frontmatter}

\section{Introduction}

It is well known that  fully
incompressible materials do  not  exist in nature. However, it is important to have a   mathematical model for an incompressible
medium, as an idealization of media that exhibit extreme resistance to volume change. In the context of isothermal mechanics, an ideal incompressible material is
a medium that can only be deformed without any change in volume. Extensive literature
has been devoted to   qualitative analysis and
numerical methods for constructing solutions of incompressible fluids as limits
of compressible ones as the Mach number tends to zero (see for example
\cite{klm,LiMa}).\newline
However, when the process is not isothermal, the notion of incompressibility is
not well defined. Several possibilities arise.

For compressible fluids, the pressure is a constitutive function while for
incompressible fluids  the pressure is only a Lagrange multiplier
associated with the constraint of incompressibility. Therefore,
to compare the solutions of compressible and incompressible media, it
is convenient to choose the pressure $p$ (instead of the density $\rho$) and the
temperature $T$ as thermodynamic  variables (see, e.g. \cite{Raj,Raj1}),
the other quantities, such as specific volume $V=1/\rho$ and internal energy $%
\varepsilon$, being determined by constitutive equations in the form:
\begin{equation*}
V \equiv V (p,T)\,,\quad \varepsilon \equiv \varepsilon (p,T).
\label{const1}
\end{equation*}
Two parameters are important for a fluid: the \emph{thermal expansion coefficient} $%
\alpha$ and the \emph{compressibility factor} $\beta$ defined by
\begin{equation}
\alpha = \frac{V_T}{V},\qquad \beta = -\frac{V_p}{V}\,,  \label{parameters}
\end{equation}
where the subscripts $T$ and $p$  indicate partial derivatives with respect to variables $%
T$ and $p$. \newline
Experiments confirm that  for fluids considered as incompressible  the volume
changes little with the temperature and remains practically unchanged  with the
pressure. For this reason, many authors consider as incompressible a
material for which the specific volume does not vary with   the pressure but varies only with the temperature (\emph{i.e.}, $V \equiv V(T)$).\newline

The first model of incompressibility was proposed by M\"{u}ller \cite%
{Muller1}: Here, all the constitutive equations of an incompressible fluid
do not depend on the pressure. We think that M\"{u}ller's motivation stems
from his using of variables $\rho$ and $T$: in experiments, the density of
incompressible materials depends on the temperature and it is reasonable to
assume that this is the case for all constitutive functions as for instance for  the internal energy. \newline
Nevertheless, M\"{u}ller proves that the only function $V \equiv V(T)$
compatible with the entropy principle is a constant. As previously
indicated, this result obviously disagrees with experimental or theoretical
results as in the so-called Boussinesq approximation (see, \emph{e.g.} \cite%
{Chandra}, \cite{brian}). We called this contradiction the \emph{M\"{u}ller
paradox} \cite{GMR}.

A second, less restrictive, model usually employed in the literature
requires that the only constitutive function independent of the pressure is
the specific volume (see, for example Rajagopal \emph{et al} \cite%
{Raj,Raj1}) and as a result the Gibbs equation is satisfied. In a recent
paper \cite{GMR}, we named such a material a \emph{%
quasi-thermal-incompressible medium} and   proved that for a pressure
smaller than a critical value, the ideal medium of M\"{u}ller can be recovered
as a limit case.\newline
Nonetheless, a weakness of   $V\equiv V(T)$ as definition for
incompressibility was first noted by Manacorda \cite{Manacorda} who showed
that instabilities occur in wave propagation. The instabilities are due to
the  non-concavity of the chemical potential and to   the sound velocity
$c$ becoming imaginary; the mathematical system of Euler equations is of
elliptic type. Let us also note that in the solid case, Dunwoody and Odgen
\cite{Dunwoody} showed that conditions for infinitesimal stability are
unattainable if the trace of the strain tensor is dependent on temperature.
Starting from Manacorda's observation, other authors such as Scott \emph{et al} \cite{Scott1,Scott2} proposed an alternative definition of
incompressibility: instead of $V\equiv V(T)$, they assumed $V\equiv V(S)$,
where $S$ is the specific entropy; then, in the ideal incompressible case,
the Mach number becomes unbounded and the sound velocity becomes infinite.
From an experimental point of view, this assumption is unrealistic because
the entropy is not an observable and no direct evaluation of $V\equiv V(S)$
is possible. Moreover, this ideal limit case implies a parabolic structure
for the Euler equations.

Due to the above considerations, the aim of our paper is to propose a
more realistic model of incompressible medium as limit case of
 a quasi-thermal-incompressible material based on the following requirements:\newline
\emph{(i)} \ The model must respect all the principles of thermodynamics:
both the Gibbs equation and the thermodynamic  stability corresponding to
the concavity of the chemical potential must be satisfied.\newline
\emph{(ii)}\ The model must fit with experiments when the compressibility
factor is not zero but it is very small. \newline
It is noteworthy that we obtain numerical results perfectly fitting with
those obtained in paper \cite{GMR}. The results are also extended to
hyperelastic materials.

\section{Thermodynamic restrictions}

We consider the thermodynamic  conditions verified by compressible fluids
when the specific volume is governed by a constitutive function written in the
form:
\begin{equation}
V\equiv V(p,T).  \label{VpT}
\end{equation}
The entropy principle and the thermodynamic stability must be satisfied.

$\qquad $ a) \textbf{\emph{The entropy principle}}

\noindent In local equilibrium, the entropy principle requires the validity
of the Gibbs equation:
\begin{equation}
TdS=d\varepsilon +p\, dV .  \label{Gibbs}
\end{equation}
In addition for Navier-Stokes-Fourier fluids, the heat conductivity
and the viscosity coefficients   must be non negative  although
they are null for Euler fluids. The choice of independent variables $p$\,
and $T$ induces the chemical potential $\mu$ as a natural thermodynamic
potential:
\begin{equation*}
\mu =\varepsilon +p\,V-T\,S  \label{chemical potential}
\end{equation*}
and Eq. (\ref{Gibbs}) is equivalent to
\begin{equation*}
d\mu =V dp - S dT.  \label{diff che pot}
\end{equation*}
Other thermodynamic variables derive from the chemical potential:
\begin{equation}
V =\mu_p \, , \qquad S = - \mu_T  \label{Gibbs relations}
\end{equation}
and the specific internal energy is
\begin{equation}
\varepsilon= \mu- p \ \mu_p -T\ \mu_T ,  \label{internal energy}
\end{equation}
where for any function $f\equiv f(p,T)$, we denote
\begin{equation*}
f_p = \left(\frac{\partial f}{\partial p}\right)_T, \quad f_T = \left(\frac{%
\partial f}{\partial T}\right)_p.
\end{equation*}
The thermal equation of state (\ref{VpT}) is determined by experiments
while the chemical potential and the entropy density can be deduced as
follows:
\begin{equation}
\mu= \int V(p,T)\, dp + \widetilde{\mu}\,(T), \qquad S= -\int V_T(p,T)\, dp
- \widetilde{\mu}^\prime\,(T),  \label{munew}
\end{equation}
where $\widetilde{\mu}\,(T)$ is a  function only depending on $T$.
From Eq. (\ref{internal energy}) we get
\begin{equation}
\varepsilon= e(T) + \int V \, dp -p\,V -T\int V_{_T} \, dp,
\label{internal energy 2}
\end{equation}
with
\begin{equation}
e(T) \equiv\, \widetilde{\mu}\,(T)-T\,\widetilde{\mu}^{\ \prime }(T) \quad
\text{or equivalently} \quad \widetilde{\mu}(T) = -T \int{\frac{e(T)}{T^2}\,
dT}.  \label{emutilde}
\end{equation}
We summarize this as a statement:

\begin{stat}
- For any constitutive functions $V\equiv V(p,T)$ and $e\equiv e(T)$, the
entropy principle is satisfied if the chemical potential, entropy density
and internal energy are given by Eq. \emph{(\ref{munew})}$_1$, Eq. \emph{(%
\ref{munew})}$_2$ and \emph{(\ref{internal energy 2})}, together with Eq.
\emph{(\ref{emutilde})}$_2$.
\end{stat}

\smallskip

$\qquad$ b) \textbf{\emph{Thermodynamic  stability}}

\noindent The specific heat {\footnotesize C}{$\rm _{^{_p}}$} is defined as the partial derivative of
the specific enthalpy $h \equiv \varepsilon + p\ V$ with respect to $T$ at
constant pressure $p$. Consequently, Eq. (\ref{internal energy 2}) yields
\begin{equation}
 {\rm \footnotesize C}{\rm _{^{_p}}}  = e^{\prime }(T) - T\displaystyle \int V_{TT}\, dp \ .  \label{CP}
\end{equation}
Thermodynamic stability requires that the chemical potential be a
concave function of $p$ and $T$:
\begin{equation}
\mu_{{pp}}= V_p <0, \qquad \left( \mu_{_{TT}}= -\frac{{\rm \footnotesize C}{\rm _{^{_p}}}}{T} <0 \right)
\label{muTT}
\end{equation}
and
\begin{equation}
I \equiv \mu_{_{TT}}\,\mu_{{pp}}- \mu_{_{T}p}^2 = -\frac{{\rm \footnotesize C}{\rm _{^{_p}}}\,V_p}{T}%
-V_T^2>0 \quad \Longleftrightarrow \quad V_p < -\frac{T\, V_T^2}{{\rm \footnotesize C}{\rm _{^{_p}}}}.
\label{I}
\end{equation}
By using Eq. (\ref{parameters}), inequality (\ref{I}) can be written in
terms of the thermal expansion coefficient $\alpha$ and the compressibility factor $\beta$:
\begin{equation}
\beta > \beta_{cr}, \qquad \beta_{cr}= \frac{\alpha^2 T V}{{\rm \footnotesize C}{\rm _{^{_p}}}} >0.
\label{betacr}
\end{equation}

\begin{stat}
- Thermodynamic stability requires that the state functions $V\equiv V(p,T)
$ and $e\equiv e(T)$ satisfy the inequalities:
\begin{equation*}
V_p < -\frac{T\, V_T^2}{{\rm \footnotesize C}{\rm _{^{_p}}}}, \qquad {\rm \footnotesize C}{\rm _{^{_p}}} >0 .  \label{V_p}
\end{equation*}
Consequently, there exists a lower bound limit $\beta_{cr}$ of $\beta$ such
that if $\beta >\beta_{cr}$, then the material is stable.
\end{stat}

\noindent The adiabatic sound velocity $c$ is:
\begin{equation*}
c^{2}=\left( \frac{\partial p}{\partial \rho }\right) _{S}=-V^{2}\left(
\frac{\partial p}{\partial V}\right) _{S}.
\end{equation*}
From Eq. (\ref{Gibbs relations}) we obtain
\begin{equation}
dV=\mu _{pT}\,dT+\mu _{pp}\,dp, \qquad dS=-\mu _{TT}\,dT-\mu _{Tp}\,dp.\
\label{calcoloc}
\end{equation}
When $dS=0$, Eq. (\ref{calcoloc})$_2$ substituted in Eq. (\ref{calcoloc})$_1$
yields $p$ as a function of $V$ and we get:
\begin{equation}
c^{2}=-\frac{\mu _{p}^{2}\ \mu _{TT}}{I}.  \label{sound velocity}
\end{equation}
Therefore, when the chemical potential is a concave function of $p$ and $T$,
we automatically get $c^{2}>0$ and the differential system for Euler fluids
is hyperbolic. Taking account of Eqs. (\ref{munew})$_1$, (\ref{emutilde})$_2$%
, (\ref{CP}), (\ref{I}) and (\ref{betacr}),  Eq. (\ref{sound velocity})
yields the sound velocity in the form:
\begin{equation*}
c=\sqrt{\frac{V}{\beta-\beta_{cr}}}\, .  \label{sound-vel}
\end{equation*}
We observe that when $V\equiv V(T)$ the model of incompressibility
corresponds to $\beta=0$ and for Euler fluids the differential system is
elliptic; when $V\equiv V(S)$, the model of incompressibility corresponds to $%
\beta\equiv \beta_{cr}$ and the system  is parabolic. In our case $%
\beta>\beta_{cr}$, the differential system is hyperbolic and the fluid is
stable.

\section{Quasi-Thermal-Incompressible Materials}

For the so-called \emph{incompressible fluids}, the volume changes little
with the temperature and changes very little with the pressure. Experiments
confirm this assumption.\newline
In the neighborhood of a reference state $(p_{_0},T_{_0},V_{_0})$, we choose
a small dimensionless parameter $\delta \ (\delta \ll 1) $ such that:
\begin{equation}
\delta= \alpha_{_0}\, T_{_0},  \label{delta}
\end{equation}
and moreover, we assume that $\beta_{_0}$ is of order $\delta^{2}$:
\begin{equation}
\beta_{_0} \, p_{_0} = O(\delta^2),  \label{beta}
\end{equation}
where $\alpha_{_0}$ and $\beta_{_0}$ are the thermal expansion coefficient and the
compressibility factor at the reference state.

\begin{definition}
- {A compressible fluid satisfying the thermodynamic conditions of Section
2 is called an Extended-Quasi-Thermal-Incompressible  fluid (EQTI) if there
exist $\widehat{V}(T)$ and $\widehat{\varepsilon}(T)$ such  that}
\begin{equation}
V(p,T)= \widehat{V}(T) + O(\delta^2) \ \mathrm{with} \ \widehat{V}%
^{\prime}(T)= O(\delta) \ \mathrm{and}\ \ \varepsilon(p,T)=\widehat{%
\varepsilon}(T) + O(\delta^2) .  \label{qti}
\end{equation}
\end{definition}

This means that an \emph{EQTI} material is a stable compressible fluid that
approximates an incompressible fluid  to order  $\delta^2$ in the sense
of M\"uller's definition. \newline
Conditions (\ref{delta})-(\ref{beta}) together with Eq. (\ref{qti})$_1$
yield the representation of $V(p,T)$:
\begin{equation}
V(p,T)= V_{_0} + \delta \, W(T) - \delta^2 U(p,T),  \label{VV}
\end{equation}
where $W(T)$ and $U(p,T)$ are two constitutive functions chosen in agreement
with conditions in Section 2. From Eqs. (\ref{munew})$_1$ and (\ref{VV}) we
deduce:
\begin{equation*}
\mu= \widetilde{\mu}\,(T) + p\, V_{_0} +\delta \, p \ W(T) +\delta^2
\widehat{\mu}(p,T), \quad \widehat{\mu}(p,T)= - \int{U(p,T) \ dp}.
\label{muqti}
\end{equation*}
Since we can incorporate its limit value in $\widetilde{\mu}\,(T)$ without
loss of generality, we can chose the function $\widehat{\mu}$ so that $%
\lim_{p->0} \widehat{\mu}(p,T)=0$. From Eq. (\ref{internal energy 2}) we
obtain:
\begin{equation}
\varepsilon(p,T)= e(T) - \delta \, T\, W^\prime(T)\ p + O(\delta^2).  \label{e}
\end{equation}
Due to Eq. (\ref{e}), in order to satisfy Eqs. (\ref{qti})$_2$ we require that  the
pressure cannot exceed a critical value:
\begin{equation}
p \ll p_{cr}(T) \qquad \mathrm{with}\qquad p_{cr}(T)=\frac{1}{\delta} \frac{%
e(T)}{T\, W^\prime(T)}.  \label{pcrg}
\end{equation}
We observe that the critical value is large and of order   $\delta^{-1}$%
. In incompressible materials, the pressure cannot exceed a critical value
depending on temperature. Inequality (\ref{pcrg}) implies
\begin{equation}
\varepsilon(p,T)= e(T) + O(\delta^2).  \label{eee}
\end{equation}
Moreover, from Eqs. (\ref{parameters}) and (\ref{delta}) we obtain
\begin{equation*}
\alpha = \delta \ \frac{W^{\prime }(T)}{V_{_0}} + O(\delta^2), \quad
W^{\prime }(T_{_0})= \ \frac{V_{_0}}{T_{_0}}; \qquad \beta = \delta^2 \
\frac{U_p(p,T)}{V_{_0}}.  \label{alfabeta}
\end{equation*}
From Eqs. (\ref{CP}) and (\ref{betacr}) we get
\begin{equation*}
\beta_{cr} = \delta^2\,\frac{T\,{W^\prime}^2 (T)}{V_{_0}\,{\rm \footnotesize C}{\rm _{^{_p}}}}, \qquad
{\rm \footnotesize C}{\rm _{^{_p}}}\simeq e^{\prime }(T),
\end{equation*}
and the thermodynamic  stability is ensured when
\begin{equation}
U_p(p,T) > \frac{T\,{W^\prime}^2 (T)}{{\rm \footnotesize C}{\rm _{^{_p}}}}, \qquad {\rm \footnotesize C}{\rm _{^{_p}}} \simeq e^\prime(T)>0.
\label{newineq}
\end{equation}
Finally, from Eq. (\ref{sound velocity}) we deduce that the dominant part in
sound velocity is of order  $\delta^{-1}$:
\begin{equation}
c \simeq \frac{V_{_0}}{\delta} \sqrt{\frac{{\rm \footnotesize C}{\rm _{^{_p}}}}{U_p\,{\rm \footnotesize C}{\rm _{^{_p}}} -T {W^\prime}^2}} \
.  \label{sv}
\end{equation}
We can conclude with

\begin{stat}
- An EQTI fluid given by constitutive functions \emph{(\ref{VV})} {and}
\emph{(\ref{eee})}, satisfying inequality \emph{(\ref{newineq})} is a good
approximation of an incompressible fluid as $\ V$ and $\varepsilon$ differ to order $\delta^2$ from functions depending only on $T$, provided the
pressure is smaller than a critical pressure $p_{cr}$ {given by Eq.} \emph{(%
\ref{pcrg})}. The sound velocity given by Eq. \emph{(\ref{sv})} is real.
\end{stat}

\noindent \textbf{Remark}: We notice that in the limit case of isothermal
processes for which $W(T)\equiv 0$ and $U\equiv U(p)$, we have $\beta_{cr}
\equiv 0$ and inequality (\ref{newineq})$_1$ and the sound velocity become $%
U^\prime(p)>0$ and $c= V_0/(\delta \sqrt{U^\prime})$, respectively. In this
case again, the \emph{EQTI} requires that $V$ is not constant but is
function of $p \, $ in the form\, $V= V_0 - \delta^2 U(p)$.

\subsection{Linear dependence of $V$ with respect to $T$ and $p$}

The most significant case is the linear expansion of $\,V$ near $%
(T_{_{0}},p_{_{0}})$:
\begin{equation}
V=V_{_{0}}\{ 1\,+\alpha\, (T-T_{_{0}})-\,\beta\, (p-p_{_{0}})\, \} \quad
\text{with}\quad e={\rm \footnotesize C}{\rm _{^{_p}}}\,T\,.  \label{linV}
\end{equation}
In this case, the scalars $\alpha$, $\beta$ and {\footnotesize C}{$\rm _{^{_p}}$} are positive constants.
Expression (\ref{linV}) is a particular case of Eq. (\ref{VV}) under the
identifications:
\begin{equation}
\alpha\,T_{_0}=\delta, \quad W(T)=\frac{V_{_0}}{T_{_0}}\,(T-T_{_0}), \quad
U(p,T) = \frac{\beta}{\alpha^2}\frac{V_{_0}}{T_{_0}^2}\,(p-p_{_0}).
\label{casolineare}
\end{equation}
Then, the fluid is \emph{EQTI} if
\begin{equation}
\alpha\,T_{_0}=\delta\ll 1 \quad \text{and} \quad \beta > \beta_{cr} \quad
\text{with}\quad\beta_{cr} = \delta^2 \frac{V_{_0}}{{\rm \footnotesize C}{\rm _{^{_p}}}\,T_{_0}} ,
\label{beta3}
\end{equation}
together with
\begin{equation}
p \ll p_{cr}, \qquad \mathrm{with}\qquad p_{cr}=\frac{1}{\delta}\frac{%
{\rm \footnotesize C}{\rm _{^{_p}}}\,T_{_0}}{V_{_0}}.  \label{inequ2}
\end{equation}
Relation (\ref{sv}) and relation (\ref{casolineare}) yield the adiabatic
sound velocity:
\begin{equation}
c_{_0} = \sqrt{\frac{{\rm \footnotesize C}{\rm _{^{_p}}} V_{_0}}{ \beta \,{\rm \footnotesize C}{\rm _{^{_p}}} - \,\alpha^2 V_{_0} T_{_0}}} \qquad
\mathrm{or}\qquad \beta= \frac{V_{_{0}}}{c_{_0}^2}+ \alpha^2\, \frac{T_{_0} V_{{_0}}%
}{{\rm \footnotesize C}{\rm _{^{_p}}}}\, .  \label{c2}
\end{equation}
Consequently, a fluid can be considered as incompressible if the pressure is
smaller than a critical pressure which is of order $\alpha^{-1}$. This is
similar to conclusions in paper \cite{GMR}. In addition, to ensure the
convexity of the chemical potential, the compressibility factor must be very
small but not identically null: $\beta$ must be greater than a critical
value $\beta_{cr}$ which is of order $\alpha^2$. From Eq. (\ref{c2}), the
value of the sound velocity allows to calculate the value of $\beta$. We
finally observe that, in the Boussinesq approximation, $\beta \neq 0$
induces an additional term depending on $p$\, in the density expansion near ($%
T_{_{0}}, p_{_{0}}$):
\begin{equation*}
\rho =\rho_{_{0}}\{ 1\,-\alpha\, (T-T_{_{0}})+\,\beta\, (p-p_{_{0}})\, \} .
\label{linearV}
\end{equation*}
Therefore, these considerations should be useful to revisit and to justify
more rigorously the Boussinesq approximation. We point out that in this case
we obtain constitutive equations that may be implicit in the direction
pointed out in \cite{Raj3}.

\subsection{Case of liquid water}

Numerical values for liquid water are obtained in \cite{Handbook}:
\begin{equation*}
p_{_{0}} = 10^{5}\,\mathtt{Pascal}; \ V_{_{0}}=10^{-3}\,\mathtt{m}^3/\mathtt{%
kg}; \ T_{_{0}} = 293^{\ \circ }\mathtt{K}; \ {\rm \footnotesize C}{\rm _{^{_p}}}=4.2\times 10^{3}\mathtt{Joule}/%
\mathtt{kg}^{\ \circ}\mathtt{K};
\end{equation*}
\begin{equation*}
\ c=1420\ \mathtt{m}/ \mathtt{sec}\,; \ \alpha =2.07\times 10^{-4}/\,^{\circ
}\mathtt{K}; \ \beta _{cr}=3\times 10^{-12}/\,\mathtt{Pascal}.
\end{equation*}
Inequality (\ref{inequ2}) reads
\begin{equation*}
p \ll p_{cr} = 2 \times 10^{10}\,\mathtt{Pascal} = 2\times 10^{5}\, \mathtt{%
atm},  \label{inequ4}
\end{equation*}
which is the same critical pressure as in paper \cite{GMR}. We observe that
Eq. (\ref{c2}) yields $\beta =4.98\times 10^{-10}/\,\mathtt{Pascal} $ which
automatically satisfies the inequality (\ref{beta3}). Liquid water is indeed a
very good example of an \emph{EQTI} liquid.

\section{Quasi-thermal-incompressibility for hyperelastic media}

\subsection{Generalities}

As done in paper \cite{GMR} and by taking up the definition of pressure
presented by Flory in 1953 for rubber gum \cite{Flory2,Flory} and extended
for hyperelastic media by Gouin and Debi\`eve in 1986 \cite{Gouin}  and Rubin
in 1988 \cite{Rubin}, we can extend the   results from fluids to thermo-elastic
materials. With this aim we define
\begin{equation}
\bm{\widetilde{C}} = \frac{1}{(\det \bm{C})^{\frac{1}{3}}} \, \bm{C}\qquad
\text{or} \qquad \bm{C} = J^{\frac{2}{3}} \ \bm{\widetilde{C}}\,,  \label{ci}
\end{equation}
where $\bm{C}= \bm{F}^T\bm{F}$ is the right Cauchy stress deformation
tensor, $\bm{F}$ is the deformation gradient, $J= \det \bm{F}=\rho_{_{0}}
/\rho$ and $\rho_{_{0}}$ is the reference density. The specific free energy
can be expressed in the form:
\begin{equation*}
\psi \equiv f(\rho,\bm{\widetilde{C}},T)\,,
\end{equation*}
where the independent variables $\rho$ and $\bm{\widetilde{C}}$ are used
instead of $\bm{C}$. Since $\det \bm{\widetilde{C}}$ =1, the variable $\rho$
corresponds to the change of volume, while the tensorial variable $%
\bm{\widetilde{C}}$ is associated with the \emph{distortion} of the medium:
we call $\bm{\widetilde{C}}$ \emph{the pure deformation of the hyperelastic
body}. This point is fundamental for the decomposition of the stress tensor.
If $f$ is independent of $\bm{\widetilde{C}}$, then we are back to the fluid
case. It is convenient to introduce the function $g$ such that:
\begin{equation*}
g(\rho,\bm{C},T)\equiv f\left(\rho,\frac{1}{(\det \bm{C})^{\frac{1}{3}}} \, %
\bm{C},T\right).
\end{equation*}
Consequently, the free energy of an hyperelastic material can be defined as $%
\psi\equiv g(\rho,\bm{C},T)$ where $g$ is a homogeneous function of degree
zero with respect to $\bm{C}$. We deduce the Cauchy stress tensor of the
medium in the form \cite{Gouin}:
\begin{equation}
{\bm{t}} =-p\,\bm{1}+{{\ \bm{\tau}}}\,,\quad \mathrm{with}\quad %
\displaystyle p=\rho^2 \displaystyle\frac{\partial g}{\partial \rho }\,, \ \ %
\bm{\  {\tau}}=2\rho\, \bm{F}\frac{\partial g}{\partial \bm{C}}\,\bm{F}^{T}\
\mathrm{and}\ \ \mathrm{{tr}\, {\bm{ {\tau}}}=0 \, ,}  \label{dcp}
\end{equation}%
where $\mathrm{tr}$ is the trace operator. If $g$ is independent of $\bm{C}$%
, $p\,$ corresponds to the thermodynamic pressure. As proved in \cite%
{Gouin}, $p$\ must be considered as the \emph{pressure of the hyperelastic
material}. Let us note that the pressure and the change of volume are
observable: they can be experimentally measured by using a spherical elastic
test-apparatus submitted to isotropic stresses. \newline
The Gibbs equation in the case of elastic materials is \cite%
{Truesdell,Ruggeri}:
\begin{equation}
TdS=d\varepsilon -\frac{1}{2\rho_{_{0}}}\ \bm{S} \cdot d\bm{C} \, ,
\label{gel}
\end{equation}%
where the \emph{dot} represents the scalar product between matrices and \ $%
\bm{S}= J\ \bm{F}^{-1}\,\bm{t}\,(\bm{F}^{T})^{-1} $ is the
second Piola-Kirchhoff stress tensor. By inserting Eq. ({\ref{dcp})$_1$ into $%
\bm{S}$, Eq. ({\ref{gel}) yields the Gibbs relation:
\begin{equation}
TdS=d\varepsilon + p \ dV - J^{-\frac{2}{3}} \ \bm{\widetilde{ \tau}} \cdot d%
\bm{C},  \label{Gibbs 1}
\end{equation}%
with $\bm{\widetilde{ \tau}}$ given by:
\begin{equation*}
\bm{\widetilde{ \tau}} =\frac{1}{2\,\rho }\, J^{\frac{2}{3}}\ \bm{F}^{-1}{\bm{
{\tau }}}\bm{F} ^{-1^{T}}.
\end{equation*}
If we take account of Eq. (\ref{ci}),
\begin{equation*}
d\bm{C=} \frac{2}{3}\ \rho \ \bm{\widetilde{C}} \ dV+ J^{\frac{2}{3}} \ d%
\bm{\widetilde{C}},
\end{equation*}%
and that
\begin{equation}
\bm{\widetilde{\tau}} \cdot \bm{\widetilde{C}} = \frac{1}{2\,\rho } \bm{F}%
^{-1}{\bm{ {\tau }}}\bm{F} ^{-1^{T}} .\, \bm{C} =\frac{1}{2\,\rho }\, \mathtt{%
tr}\left(\bm{F}^{-1}{\bm{ {\tau }}} \bm{F} ^{-1^{T}} \bm{F} ^{^{T}}\bm{F}%
\right)\equiv \frac{1}{2\,\rho }\, \mathtt{tr}\,{\bm{ {\tau }}}= 0,
\label{orto}
\end{equation}
we obtain the Gibbs relation (\ref{Gibbs 1}) in the final form
\begin{equation}
TdS=d\varepsilon + p\ dV -{\bm{ {\widetilde{\tau }}}}\cdot\, d%
\bm{\widetilde{C}}\,.  \label{Gibbs2}
\end{equation}
As in the fluid case, we introduce the chemical potential $\mu=
\varepsilon+p\,V-TS- \bm{\widetilde{\tau}} \cdot \bm{\widetilde{C}}$. Thanks
to the orthogonality condition (\ref{orto}), $\mu$ takes the same form as
for fluids:
\begin{equation*}
\mu= \varepsilon+p\,V-T S.
\end{equation*}
Equation (\ref{Gibbs2}) implies
\begin{equation}
d\mu= V dp-S dT +{\bm{ {\widetilde{\tau }}}}\cdot\, d\bm{\widetilde{C}} .
\label{difmu}
\end{equation}
Consequently, the change of variables from ($V, T$) into ($p, T$) is natural
and the variable $\bm{\widetilde{C}}$ does not change. Equation (\ref{difmu})
implies:
\begin{equation}
V= \mu_{p}\,, \quad S=-\mu_{_T}\,, \quad \bm{\widetilde{\tau}}= \mu_{ \bm{%
\widetilde{C}}}\,, \quad \varepsilon = \mu-T \mu_{_T}-p\,\mu_{p} \, ,
\label{varie}
\end{equation}
with
\begin{equation*}
\mu _{p} =\left( \frac{\partial \mu}{\partial p}\right) _{T,\, \bm{\widetilde{C}%
}},\quad \mu _{T}=\left( \frac{\partial \mu}{\partial T}\right) _{p,\, \bm{%
\widetilde{C}}}, \quad \mu _{ \bm{\widetilde{C}}}=\left( \frac{\partial \mu }{%
\partial \bm{\widetilde{C}}}\right) _{p,\,T}.
\end{equation*}
Let us assume that the specific volume is written as a function of $p, T, \bm{%
\widetilde{C}}$:
\begin{equation*}
V\equiv V (p,T, \bm{\widetilde{C}}).
\end{equation*}
By integration of Eq. ($\ref{varie})_{1}$, we obtain
\begin{equation}
\mu = \int V(p,T, \bm{\widetilde{C}})\,dp+\widehat{\mu} (T, \bm{\widetilde{C}}),
\quad S = -\int V_T(p,T, \bm{\widetilde{C}})\,dp-\widehat{\mu}_T (T, \bm{%
\widetilde{C}})\, ,  \label{cpot}
\end{equation}
where $\widehat{\mu}$ is an additive function  depending only on $T$ and $\bm{%
\widetilde{C}}$. By substituting in Eq. ($\ref{varie})_{4}$, we get:
\begin{equation}
\varepsilon= e(T,\bm{\widetilde{C}})+ \int V(p,T,\bm{\widetilde{ C}})\,dp -p\,V -T
\int V_{_T} \,dp \,,  \label{intr}
\end{equation}
with the additional function
\begin{equation}
e( T,\bm{\widetilde{ C}})=\widehat{\mu} - T \widehat{\mu}_T \ \ \text {or
equivalently} \ \ \widehat{\mu}(T,\bm{\widetilde{ C}}) = -T \int{\frac{e(T,%
\bm{\widetilde{ C}})}{T^2}\, dT}+k(\bm{\widetilde{ C}}),  \label{emu}
\end{equation}
where $k(\bm{\widetilde{ C}})$ is an  arbitrary function of $\bm{\widetilde{ C}}$. The
additional function is in the same form than for fluids. Therefore }}

\begin{stat}
- For any constitutive functions $\ V\equiv V(p,T,\bm{\widetilde{ C}})$, {and} $%
e\equiv e(T,\bm{\widetilde{ C}})$, {the entropy principle is satisfied if the
chemical potential, the entropy density and the internal energy are given by}
Eqs. \emph{(\ref{cpot})}$_1$, \ \emph{(\ref{cpot})}$_2$ , \emph{(\ref{intr})}
with Eq. \emph{(\ref{emu})}$_2$, respectively.
\end{stat}

In nonlinear theories, the concavity of the entropy density may be valid
only in some domain of state variables. In particular this is the case in
nonlinear elasticity where the concavity is in contradiction with the
objectivity principle if the deformation is large \cite{Truesdell} (see also
\cite{Dafermos,RS}). The concavity of the chemical potential is valid only
for sufficient small deformations in the neighborhood of undeformed
configuration. In the decomposition of the stress tensor given by Eq. (\ref%
{dcp}), the pressure $p$\, is the main observable variable and the pure
deformation $\bm{\widetilde{ C}}$ does not affect the change of variables between
internal energy and chemical potential. Consequently, taking $\bm{\widetilde{C}}
$ constant we  conclude that inequalities (\ref{muTT}-\ref{I}) are
necessary conditions for   stability.

\subsection{Quasi-Thermal-Incompressible Solids}

In experiments, the specific volume of incompressible solids changes if the
temperature varies. In the literature it is usually assumed that $J\equiv
J(T)$ (or equivalently $V\equiv V(T)$). Therefore, it is natural to define
\emph{EQTI} solids  similarly to the definition given for fluids.

\begin{definition}
- A compressible solid satisfying thermodynamic conditions associated with the
entropy principle and stability is called an
Extended-Quasi-Thermal-Incompressible  solid if there exist $\widehat{V}(T)$
and $\widehat{\varepsilon}(T,\bm{\widetilde{ C}})$ such  that
\begin{equation*}
V(p,T,\bm{\widetilde{ C}})= \widehat{V}(T) + O(\delta^2) \ \ \mathrm{with} \ \
\widehat{V}_T= O(\delta) \quad \mathrm{and}\quad \varepsilon(p,T,\bm{\widetilde{
C}})=\widehat{\varepsilon}(T,\bm{\widetilde{ C}}) + O(\delta^2) .  \label{qti2}
\end{equation*}
\end{definition}
Assuming for solids the linear expansion  (\ref{linV}) for $V$, we derive a similar result as for fluids:
\begin{stat}
- A  thermoelastic material with constitutive equation \emph{(\ref{linV})}  is a
stable EQTI and tends to an incompressible material under the conditions
\begin{equation*}
\beta > \beta_{cr} \quad\mathrm{with}\quad\beta_{cr} = \delta^2 \frac{V_{_0}%
}{{\rm \footnotesize C}{\rm _{^{_p}}}T_{_0}},\qquad \mathrm{and} \qquad p\ll p_{cr}, \quad \mathrm{with}%
\quad p_{cr}=\frac{1}{\delta}\frac{{\rm \footnotesize C}{\rm _{^{_p}}}T_{_0}}{V_{_0}}\,,
\end{equation*}
where $V_{_0}$ and $\delta = \alpha\,T_{_0} \ll 1$ are positive constants,
while {\footnotesize C}{$\rm _{^{_p}}$} may depend on the pure deformation $\bm{\widetilde{ C}}$.
\end{stat}

\subsection{Case of pure gum rubber}

As in paper \cite{GMR}, we consider the case of rubber as example of
hyperelastic material. To verify the conditions \emph{EQTI,} we get
experimental values from the literature: numerical values for rubber are
obtained in \cite{Handbook,Handbook2}:
\begin{equation*}
p_{_{0}} = 10^{5}\,\mathtt{Pascal};\ {\rm \footnotesize C}{\rm _{^{_p}}}=1.9\times 10^{3}\mathtt{Joule}/%
\mathtt{kg}^{\ \circ}\mathtt{K};\ V_{_{0}}=1.08\times10^{-3} \mathtt{m}^3/\mathtt{kg};
\end{equation*}
\begin{equation*}
\ T_{_{0}} = 273^{\ \circ} \mathtt{K}; \ c= 54 \mathtt{m}/ \mathtt{sec};\
\alpha =7\times 10^{-3}/ ^{\ \circ}\mathtt{K};\ \beta =4.5\times 10^{-7}/\,\mathtt{%
Pascal} ;
\end{equation*}
\begin{equation*}
\beta _{cr}=7.6\times 10^{-9}/\,\mathtt{Pascal}; \ p_{cr} = 2.5\times
10^{8}\,\mathtt{Pascal} = 2.5 \times 10^{3}\, \mathtt{atm}.
\end{equation*}%
We can immediately verify that rubber is also a good example of an \emph{EQTI}
medium. Let us note that the considered values for rubber are a little
different from those in \cite{GMR}; this   is due to the fact that the sound velocity
was measured at $273^{\circ }\mathtt{K}$ and not at $%
325^{\circ }\mathtt{K}$, as in \cite{GMR}.

\section{Conclusion}

Our goal was to present a model expressing the limit case of incompressible
materials and fitting with the principles and conditions of thermodynamics.
Earlier models were  thermodynamically deficient. Here, we
present  a new model (which we call \emph{%
Extended-Quasi-Thermal-Incompressible} model) starting from two simple
requirements:\newline
\emph{(i)}\ An incompressible medium does not physically exist but it is   the limit case of a compressible medium that  verifies all
thermodynamic  conditions.\newline
\emph{(ii)}\ On  physical grounds, the volume changes little with the
temperature and does not practically change with the pressure.\newline
As result, the conditions for the pressure are the same than in   previous paper
\cite{GMR} but an additional condition associated with the compressibility
factor must be verified. We point out that a decomposition of the stress
tensor for hyperelastic media into a pressure term associated with the \emph{%
change of volume} and a null-trace part associated with the \emph{pure
deformation} of the medium leads to   analogous conditions of \emph{%
EQTI} for fluids and hyperelastic materials. Let us note that on a
phenomenological basis a theory where the elasticity parameters
depend on the pressure has been developed in \cite{Raj2,Raj4}.

From a mathematical standpoint, the difficult question of proving that
the solutions depend continuously on $\delta$ and that the limit  exists
as $\delta$ tends to zero, remains an open problem.

{\small {\textbf{Acknowledgments}: The authors are grateful to Professor Constantine Dafermos,
Professor Giuseppe Saccomandi and the  anonymous reviewers for their interest on this paper
and their useful  suggestions.}}

\end{document}